\documentclass[twocolumn]{aastex631}

\usepackage[utf8]{inputenc}
\usepackage{threeparttable}
\usepackage{amsmath}
\usepackage{epstopdf}
\usepackage{threeparttable}
\usepackage{graphicx, color, url}
\usepackage{bm}
\usepackage{soul}
\usepackage{amssymb} 
\usepackage{booktabs}

\usepackage{txfonts}

\newcommand{\harm}{\texttt{HARM}}

\newcommand{\harmpi}{\texttt{HARMPI}}

\newcommand{\eqb}{\begin{eqnarray}}
\newcommand{\eqe}{\end{eqnarray}}

\submitjournal{APJL}

\begin{document}

\title{Striped Jets in Post Neutron Star Merger Systems}

\correspondingauthor{Emma Kaufman, I.M. Christie }
\email{ekaufman@u.northwestern.edu, ichristi231@gmail.com}

\author[0000-0002-7236-284X]{Emma Kaufman}
\affiliation{Center for Interdisciplinary Exploration \& Research in Astrophysics (CIERA), Physics \& Astronomy, Northwestern \\ University, Evanston, IL 60208, USA}

\author{I.M.~Christie}
\noaffiliation
%\affiliation{Center for Interdisciplinary Exploration \& Research in Astrophysics (CIERA), Physics \& Astronomy, Northwestern \\ University, Evanston, IL 60208, USA}

\author{A. Lalakos}
\affiliation{Center for Interdisciplinary Exploration \& Research in Astrophysics (CIERA), Physics \& Astronomy, Northwestern \\ University, Evanston, IL 60208, USA}

\author{A. Tchekhovskoy}
\affiliation{Center for Interdisciplinary Exploration \& Research in Astrophysics (CIERA), Physics \& Astronomy, Northwestern \\ University, Evanston, IL 60208, USA}

\author{D.~Giannios}
\affiliation{Department of Physics and Astronomy, Purdue University, 525 Northwestern Avenue, West Lafayette, IN 47907, USA}

\begin{abstract}
Models invoking magnetic reconnection as the particle acceleration mechanism within relativistic jets often adopt a gradual energy dissipation profile within the jet. However, such a profile has yet to be reproduced in first-principles simulations.
Here, we perform a suite of 3D general relativistic magnetohydrodynamic simulations of post-neutron star merger disks with an initially purely toroidal magnetic field. We explore the variations in both the microphysics (e.g., nuclear recombination, neutrino emission) and system parameters (e.g., disk mass).
In all our simulations, we find the formation of magnetically striped jets. The stripes result from the reversals in the poloidal magnetic flux polarity generated in the accretion disk. The simulations display large variations in the distributions of stripe duration, $\tau$, and power, $\langle P_{\Phi} \rangle$. 
We find that more massive disks produce more powerful stripes, the most powerful of which reaches $\langle P_{\Phi} \rangle \sim 10^{49}$~erg~s$^{-1}$ at $\tau \sim 20$~ms. 
The power and variability that result from the magnetic reconnection of the stripes agree with those inferred in short duration gamma-ray bursts.
We find that the dissipation profile of the cumulative energy is roughly a power-law in both radial distance, $z$, and $\tau$, with the slope in the range, $\sim 1.7-3$; more massive disks display larger slopes.
\end{abstract}


\section{Introduction} \label{sec:intro}

Relativistic jets are ubiquitous in many black hole (BH) accreting systems, such as supermassive BHs in active galactic nuclei (AGN) to core-collapse and neutron-star (NS) merger gamma-ray bursts (GRB; long and short, respectively). 
They are powerful gamma-ray emitters that often display large variability in their emission. 
However, despite the detection of several thousands of long GRBs and hundreds of short GRBs \citep[for catalog, see][]{fong2015}, the emission processes and mechanisms for particle acceleration responsible for the prompt emission remain elusive. 
In the former, we refer to the long-debated synchrotron \citep{katz1994,sari1996} vs. photospheric \citep{goodman1986,Thompson1994,meszaros_rees_2000} origin of the prompt emission.  
In this Letter, we focus on the latter: the mechanism for particle acceleration within GRB jets. 

Such a mechanism has to depend upon how the jets are launched. 
For hydrodynamic flows \citep[for recent 3D simulations, see e.g.][]{Gottlieb2020}, internal shocks and mixing can be generated while the jet breaks out of the surrounding medium, i.e., the stellar envelope for long GRBs and dynamical ejecta for short GRBs, at which point particles can be heated to mildly relativistic temperatures \citep{rees_1994,lazzati_2010}.
However, relativistic jets from BH systems have been long argued to be highly magnetized and magnetically driven \citep{blandford_electromagnetic_1977} and are, therefore, more susceptible to dissipate via magnetic reconnection rather than internal shocks \citep[see][]{sironi_petropoulou_giannio2015}. 
Magnetic reconnection is an efficient process for dissipating the jet's magnetic energy \citep{Spruit2001,Drenkhahn_spruit_2002} to particle acceleration and has been adopted in many recent studies for reproducing the prompt emission of GRBs \citep{giannios_2008,beniamini_granot_2016,Beniamini_2017,begue2017}. 

This raises an important question: how is reconnection triggered within the jet?
Recent 3D general relativistic magnetohydrodynamic (GRMHD) simulations show that magnetized jets can profusely dissipate their magnetic energy due to magnetic instabilities both in short \citep{oreshortgrb} and long \citep{orelonggrb} GRB contexts. However, the instabilities and the associated reconnection tend to develop when the jets run into the ambient medium. An alternate mechanism is for
the the wound-up magnetic field lines in a jet to develop a ``striped wind'' configuration, in which their polarity flips over a characteristic timescale, generated by a
large-scale poloidal (i.e.pointing in R- and z-directions) dynamo within the surrounding BH accretion disk \citep[e.g magneto rotational instability, MRI, see][]{balbus_hawley_1991}.
Although this striped jet scenario has been utilized in theoretical models \citep{Drenkhahn2002,Drenkhahn_spruit_2002,Giannios2019}, 
the quest of finding it in simulations of GRB central engines and their jets has remained elusive.
The requirement for large-scale poloidal magnetic flux in launching relativistic jets \citep{blandford_electromagnetic_1977} has often been satisfied in previous studies embedding GRB accretion disks with purely poloidal magnetic fields \citep[][]{fernandez2018,Christie2019}. 
Doing so, however, allows for the accumulation of substantial poloidal flux on the BH to occur on short timescales, thus negating the effect of any small scale loops of opposite polarity created under the dynamo. However, \cite{Christie2019} demonstrated for the first time, using 3D GRMHD simulations, that an initially toroidal magnetic field embedded within a post NS merger remnant disk can generate large-scale poloidal magnetic flux of alternating polarity, launching tightly collimated relativistic striped jets\footnote{\cite{Liska_dynamo} similarly found, for radially extended accretion disks, the accumulation of large-scale poloidal flux through the black hole and relativistic jets (see Discussion section for comparisons).}.

These recent results trigger a series of important questions. Under what conditions are the striped jets produced? What are the properties of the magnetic flux reversals within the disk and jets? Assuming the magnetic polarity reversals between the stripes trigger reconnection events in the jets, what is the distribution of energy dissipation along the jets?
Here, we investigate these questions by performing 3D GRMHD simulations of  post NS merger remnant disks, with initially purely toroidal magnetic fields and a variety of initial conditions. In Sec.~\ref{sec:setup}, we discuss the numerical setup of our simulations. In Sec.~\ref{sec:results}, we discuss the properties of magnetic flux reversals within the disk and jets.
In Sec.~\ref{sec:discussion}, we conclude.

\section{Numerical Setup}
\label{sec:setup}

\begin{table}

\centering
\hspace*{-\leftmargin}\begin{threeparttable}
\centering
\begin{tabular}{@{}c|c|cc|c@{}}
\hline
Model & Total resolution & \multicolumn{2}{c|}{Duration, $t_{\rm max}$} & Initial Torus\\
Name & $N_r\times N_\theta \times N_\phi $ & $(s)$ & $(10^5  \, r_g/c)$ & Mass $(10^{-2} \, M_\odot)$\\
\hline
BT & $512 \times 256 \times 128$ & $4.3$ & $2.9$ & $3.3$\\
\hline
BT-NR & $512 \times 256 \times 128$ & $0.67$ & $0.45$ & $3.3$\\
\hline
BT-LD  & $512 \times 256 \times 256$ &  $0.52$ & $0.35$ & $10$\\
\hline
BT-LDN & $512 \times 256 \times 256$ & $0.19$ & $0.13$ & $10$\\
\hline
\end{tabular}
\end{threeparttable}
\caption{Simulation setup for our toroidal post-merger models. Simulations BT and BT-LD include nuclear physics, while BT-NR neglects nuclear recombination and BT-LDN neglects all nuclear physics.}
\label{table:model_setups}
\end{table}

A detailed description of our simulation setup can be found in \citep[][their BT\footnote{For videos displaying the evolution of the BT model, see: \url{https://goo.gl/ct7Htx}.} model]{Christie2019}, but will be briefly summarized here. We performed our post-merger simulations using \harmpi \citep{harmpy}{}\footnote{\url{https://github.com/atchekho/harmpi}}, an enhanced version of the serial open-source code \harm{} \citep{gammie2003,noble2006}, with the addition of several physical processes, including neutrino cooling and nuclear recombination processes \citep[for more details, see][]{fernandez2018}. All simulations follow the setup of the BT model \citep{Christie2019}, namely a BH with mass $M_{\rm BH} = 3 \, M_\odot$ and dimensionless spin parameter $a = 0.8$, surrounded by a magnetized gas torus, with the initial electron fraction $Y_e = 0.1$ (see Table~\ref{table:model_setups}).
In addition to the BT model with torus mass of $0.033 \, M_\odot$, we consider three post-merger disks that have slight variations in their initial conditions (Table~\ref{table:model_setups} gives model names and parameters), namely: (i) neglecting nuclear recombination with torus mass of $0.033 \, M_\odot$ (BT-NR model), (ii) larger torus with mass $0.1 \, M_\odot$ with the inclusion of all nuclear physics of the BT model (BT-LD model), and (iii) larger torus with mass $0.1 \, M_\odot$ and neglecting all nuclear physics (BT-LDN model\footnote{For videos displaying the evolution of the mass and current density in the BT-LDN model, see: \url{https://rb.gy/uojnpk}.}). All tori were embedded with a purely toroidal magnetic field with plasma-$\beta$, i.e. ratio of gas to magnetic pressure, of $5$, corresponding to a maximum magnetic field strength of $\sim 5 \times 10^{14}$~G within the disk for the BT and BT-NR models. For more massive tori, the magnetic field will likely be stronger. All simulations were performed beyond $0.1$~s, allowing for runtimes comparable to the typical duration ($\lesssim$ 2s) of short GRBs. 
%%%%%% FIGURE BEGIN %%%%%%%%
\begin{figure}
\centering
\includegraphics[height=0.36\textwidth]{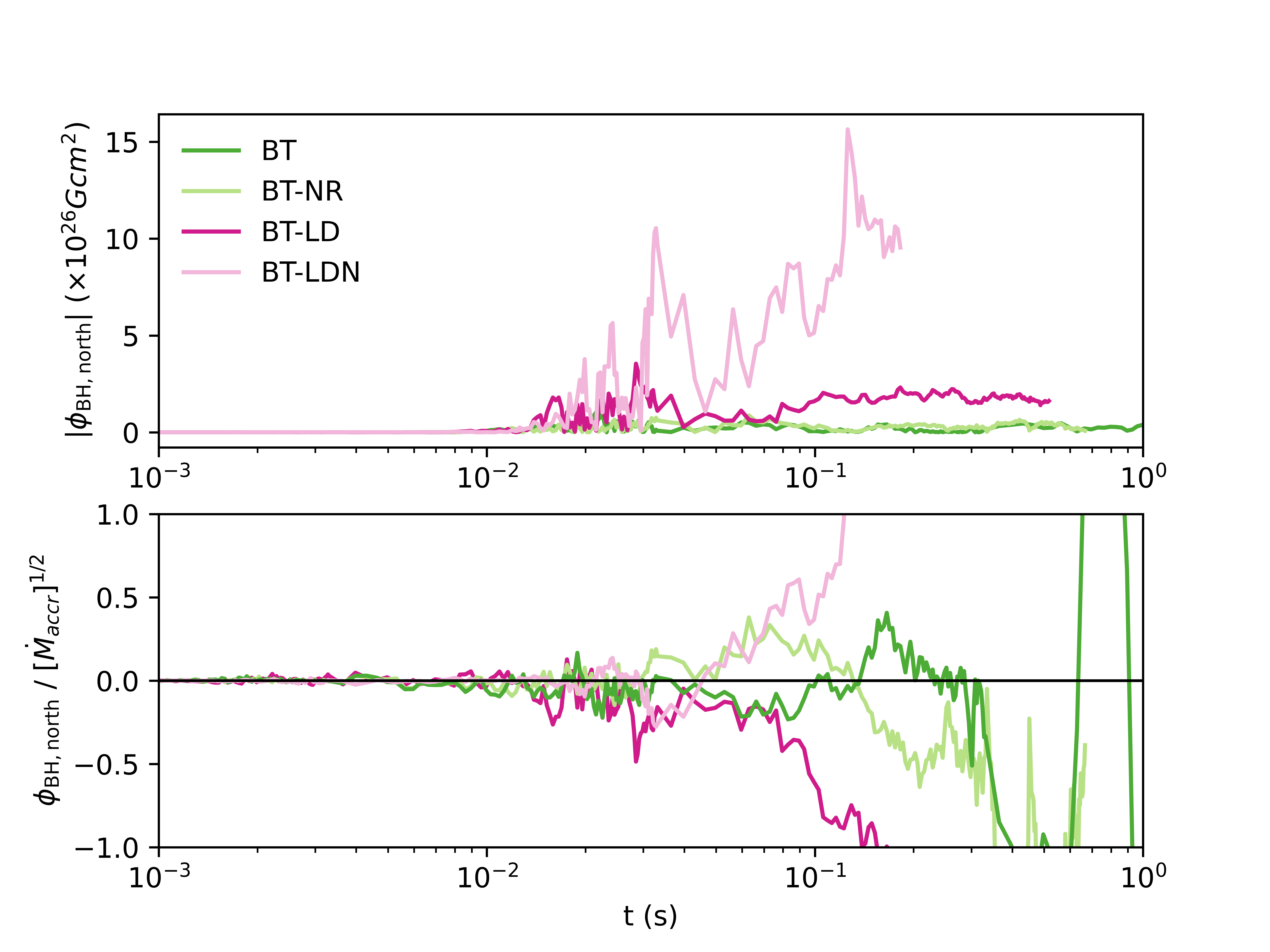}
\caption{The dynamo generates large-scale poloidal magnetic flux from a purely toroidal magnetic flux configuration and accretes it on the BH (top panel, displaying the absolute value of magnetic flux in the northern hemisphere of the BH). Additionally, the dynamo continuously generates flips in the magnetic flux (bottom panel, displaying magnetic flux normalized to the square root of the mass accretion rate). These flips in polarity result in the production of current sheets on the BH, and hence striped jets. Interestingly, neglecting all nuclear physics and neutrino emission (BT-LDN model, see legend) results in stronger BH magnetic flux, as shown in the top panel, than all other models. However, the normalized flux is comparable across all models.}
\label{fig:flux_vs_t_zoom}
\end{figure}
%%%%%% FIGURE END %%%%%%%%

\section{Results}
\label{sec:results}

Within the first $\sim 20$~ms in all our simulations, the mass accretion rate on the BH peaks \citep[see Fig.~1a in][]{Christie2019}, and launches sub-relativistic outflows which we refer to as disk winds. Interestingly, these purely toroidal models result, due to a dynamo-like action \citep[e.g. $\alpha-\Omega$ dynamo, see][]{Liska_dynamo, Christie2019}, in the development of large-scale poloidal magnetic flux on the BH (the top panel of Fig.~\ref{fig:flux_vs_t_zoom} shows the time dependance of the absolute value of the magnetic flux through the BH). Our models also launch tightly collimated relativistic jets with opening angles $\theta_{\rm j} \lesssim 5^\circ$ for BT model. \cite{Christie2019} showed that jets in the BT model were not only intermittent, but continuously flipped their magnetic polarity (i.e. resulted in striped jets), with current sheets separating the stripes of opposite polarity. This is illustrated in the bottom panel of Fig.~\ref{fig:flux_vs_t_zoom} which shows the northern component of the poloidal magnetic flux on the BH, defined as $\Phi_{\rm BH,north} = \int_{r = r_H} B^r {\rm d}A$, 
normalized to the square root of the mass accretion rate on the BH. Here, $B^r$ is the radial component of the magnetic field, $r_H = r_g [1 + \sqrt{1 - a^2}]$ is the event horizon radius, $r_g = G M_{\rm BH} / c^2$ is the gravitational radius, ${\rm d}A = \sqrt{-g} {\rm d} \theta {\rm d} \phi$ is the area element, and $g$ is the determinant of the metric.
Throughout we adopt units such that $G=M_{BH}=c=1$.
Below, we explore whether the initial conditions (more massive disk) or physical processes (nuclear physics, nuclear recombination) within the disk affect the development of current sheets within the jets.
\subsection{Properties of Magnetic Stripes}
\label{sec:properties}

As discussed in \citet{Christie2019}, the dynamo within the disk continuously generates large-scale poloidal magnetic flux of alternating polarity. This can be seen for all models in Fig.~\ref{fig:flux_vs_t_zoom} via the northern component of the magnetic flux $\phi_{\rm BH, north}$ (top panel) and normalized to the square root of the mass accretion rate on the BH (bottom panel). At early times, $t\lesssim 10^{-2}$~s, the magnitude of $\phi_{\rm BH, north}$ slowly increases while fluctuating around zero, with the BT-LD model having slightly larger flux on the BH due to the higher mass accretion rate and therefore larger accumulation of magnetic flux on the BH. However, the BT-LDN model, which neglects all nuclear and neutrino physics, displays a large accumulation of a single polarity normalized flux at 0.05 s - 0.2 s, beyond which field reversals cease.

\begin{figure}
\centering
\includegraphics[height=0.34\textwidth]{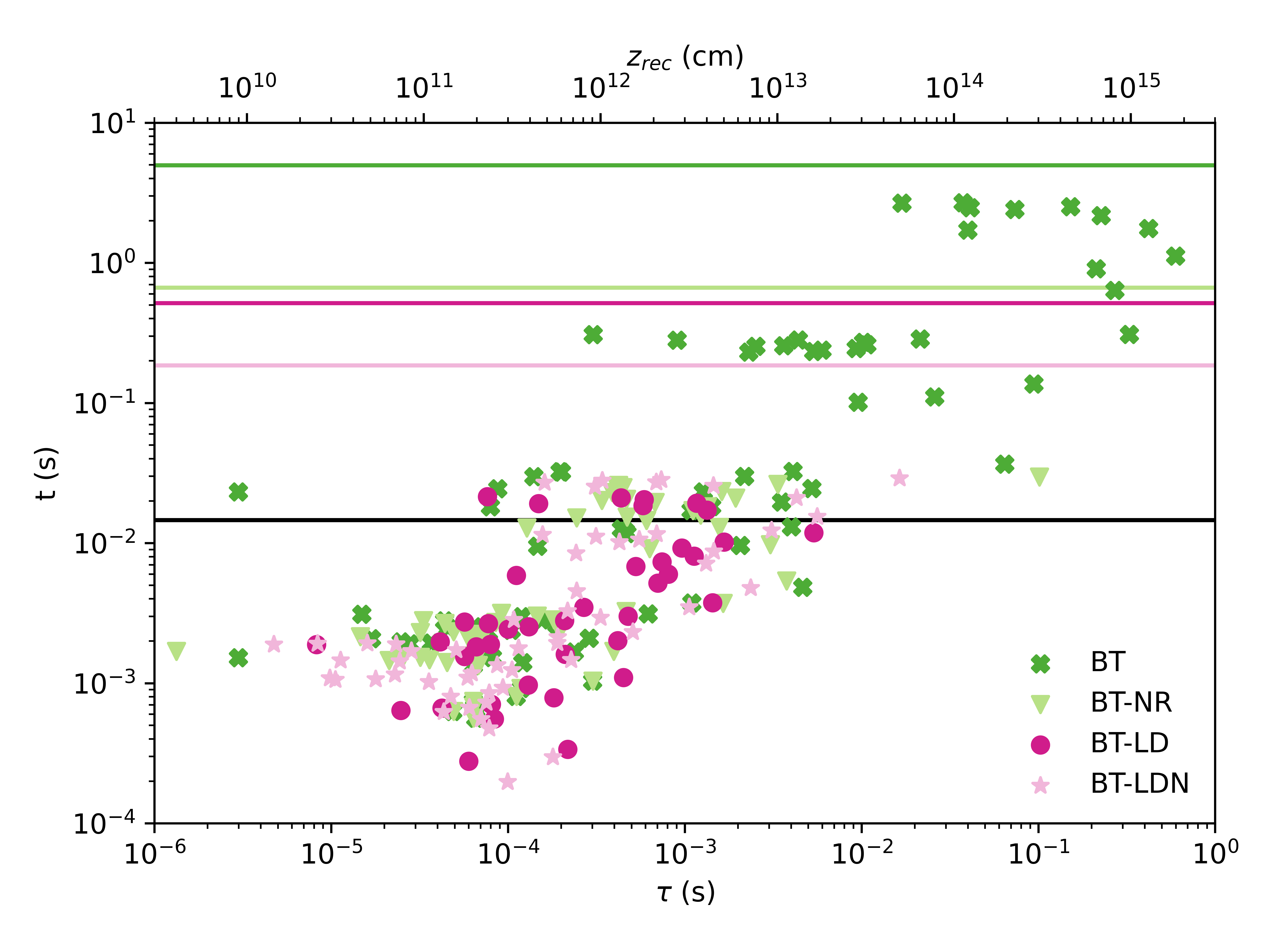}
\caption{The duration $\tau$ of field reversals on the BH, determined from $\phi_{\rm BH,north}$ (see Fig.~\ref{fig:flux_vs_t_zoom}) for all models considered (see legend), shows a general increase with increasing simulation time. For times $t \gtrsim 5\times 10^{-2}$~s, our BT-LD, BT-LDN, and BT-NR models no longer exhibit flips in polarity up to the cessation of the simulation, denoted by the horizontal lines of the same color. Reversals produced at times $\gtrsim 0.015$~s, i.e. times at which polar outflows begin to form (horizontal black line), will propagate and reconnect at various distances $z_{\rm rec}$ with the jet (see top axis). Here, we use a fidicual value of $\Gamma_{\rm j} = 100$ for the jet Lorentz factor in computing $z_{\rm rec}$ (see eqn.~\ref{eqn:z_rec}).}
\label{fig:time_vs_dt}
\end{figure}

To better understand the distribution of alternating polarity stripes seen in all of our models, we investigate the distribution of the time interval $\tau$ between the flips.
Fig.~\ref{fig:time_vs_dt} shows simulation time t as a function of stripe duration $\tau$. At early times ($10^{-4} \, {\rm s} \lesssim t \lesssim  10^{-2} \, {\rm s}$), 
all models show a large spread in duration $\tau$, which extends from $\tau \sim  10^{-5}$~s to $10^{-2}$~s. Small duration flips can occur due to small-scale poloidal magneitc flux loops, generated within the inner parts of the disk, which quickly accrete on the BH. At late times, $t \gtrsim 0.05$~s, the magnitude of $\phi_{\rm BH,north}$ increases, due to the increase in mass accretion rate and the accumulation of a single polarity flux on the BH, such that flips occur on longer timescales (i.e. $\tau \gtrsim 5 \times 10^{-3}$~s). Interestingly, by $t\sim 0.1$~s, the BT-LD, BT-LDN, and BT-NR models no longer exhibit polarity flips (see Fig.~\ref{fig:time_vs_dt}) and instead have accumulated enough flux of a single polarity, as shown in the bottom panel of Fig.~\ref{fig:flux_vs_t_zoom}. The BT model, however, continues to generate flux flips beyond $1$~s, resulting in longer stripe durations of $\tau \gtrsim 0.1$~s. 

%%%%%% FIGURE BEGIN %%%%%%%%
%%%%%% FIGURE END %%%%%%%%
The magnetic polarity flips can be associated with stripes propagating through the jet as long as they are produced at times after polar outflows begin to form ($t\sim 0.015$~s, denoted as the black line in Fig.~\ref{fig:time_vs_dt}).
%These stripes, which can result in the formation of current sheets on the BH and in the jets, also have an associated width $l$. 
If the polarity through the BH is fixed for time $\tau$, the resulting jetted outflow (which flies out at near the speed of light $c$) maintains its polarity over the corresponding radial scale, $l=c\tau$, as measured in the BH frame. 
Each stripe, propagating through the jet, has a comoving width $\Gamma_{\rm j} l$, where $\Gamma_{\rm j}$ is the jet's Lorentz factor. The stripes can ultimately reconnect within the jet at a radial distance
%\footnote{}
, as measured in the BH frame \citep{Giannios2019}
\eqb
\label{eqn:z_rec}
z_{\rm rec} \approx \beta_{\rm rec}^{-1} \, \Gamma_{\rm j}^{2} l.
\eqe
Here, $\beta_{\rm rec}=v_{\rm rec}/c$ is the reconnection rate, which has been estimated via first-principle particle-in-cell simulations (PIC) for varying plasma magnetizations and compositions \citep{Sironi_Spitkovsky_2014,guo2015,sgp16} and through analytical estimates \citep{Lyubarsky2005} as $\beta_{\rm rec} \approx 0.1$. The $\Gamma^2_{\rm j}$ factor in eq. \ref{eqn:z_rec} results from the combination of two relativistic effects: (i) the fact that the stripe width is a factor $\Gamma_j$ longer in the rest frame of the jet (where the magnetic reconnection takes place) and (ii) the time dilation effect that prolongs, by another factor $\Gamma_j$, how fast the reconnection takes place in the rest frame of the source. Adopting a typical GRB jet Lorentz factor of $\Gamma_{\rm j} = 100$, the stripes in our models can reconnect at distances $z_{\rm rec} \sim 10^{10}$~cm $ - 2 \times 10^{15}$~cm = $2.3 \times 10^{4} r_g - 4.5 \times  10^9 r_g$. We show the distances 
at which the dissipation within the jet takes place along the top axis of Fig.~\ref{fig:time_vs_dt}; we discuss this further in Sec.~\ref{sec:energy_dissipation}.

%%%%%% FIGURE BEGIN %%%%%%%%
\begin{figure}
\centering
\includegraphics[height=0.35\textwidth]{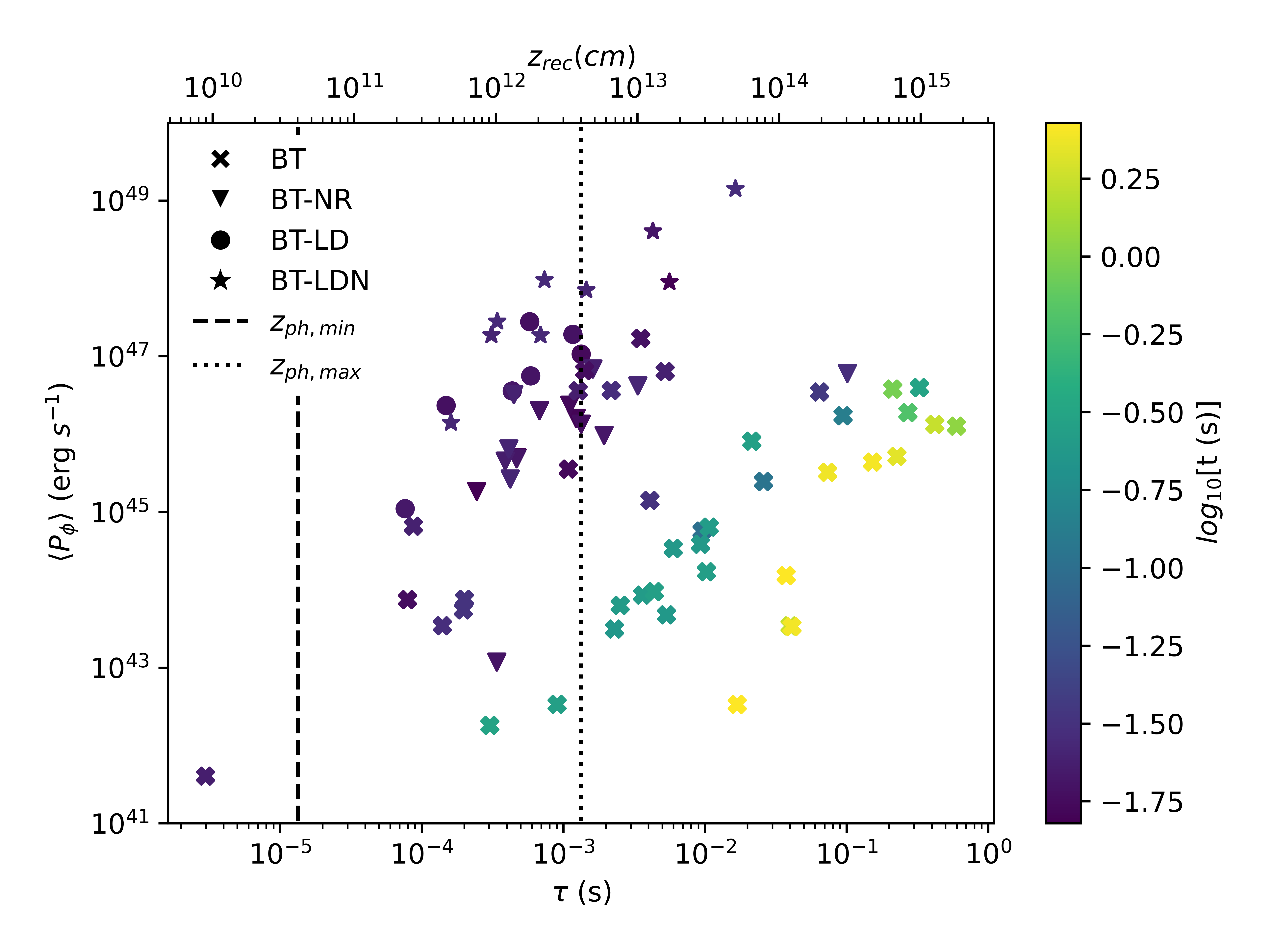}
\caption{Time-average power $\langle P_{\Phi} \rangle$ (see eqn.~\ref{eqn:p_stripe}) of each stripe propagating within the jet (see also Fig.~\ref{fig:time_vs_dt}). $\langle P_{\Phi} \rangle$ is highly scattered for all models (see legend) as a function $\tau$, with the BT model displaying additional fluctuations as a function of simulation time (see colorbar). The BT-LDN model, which neglects all nuclear physics, contains those stripes with the largest power. The black dashed (dotted) line shows the minimum (maximum) photospheric radius (see eqn.~\ref{eqn:photosphere_radius}) using fiducial values of $\Gamma_{\infty} = 300$ and $\Gamma_j = 100$.}
\label{fig:power_vs_dt}
\end{figure}
%%%%%% FIGURE END %%%%%%%%

As each stripe reconnects within the jet, it can dissipate a fraction of its magnetic energy, via relativistic magnetic reconnection, heating the plasma and accelerating particles. 
To determine the amount of dissipated energy, we can first calculate the time-averaged power within each stripe \citep{Tchekhovskoy2015},
\eqb
\label{eqn:p_stripe}
\langle P_{\Phi} \rangle \approx \frac{\omega_H^2 \, c}{96 \, \pi^2 \, r_g^2 \, \tau} \, \int_{\rm stripe} {\rm d}t \, |\Phi_{\rm BH}|^2 ,
\eqe
where $\omega_H = a/(1 + \sqrt{1 - a^2})$. Our results for $\langle P_{\Phi} \rangle$, for all models as a function of $\tau$ and time $t$ (indicated by the colorbar), are shown in Fig.~\ref{fig:power_vs_dt}.

As time increases beyond the time of initial jet formation, more magnetic flux accumulates on the BH and results in stripes of higher power. 
For a stripe of given duration $\tau$, the BT-LD and BT-LDN models generally contain stripes with larger power than the BT and BT-NR models, a direct result of the former models having a higher $\Phi_{\rm BH}$ at any given time. 

\subsection{Distribution of Energy Dissipation}
\label{sec:energy_dissipation}

%%%%%% FIGURE BEGIN %%%%%%%%
\begin{figure}
\centering
\includegraphics[height=0.39\textwidth, trim={0 5 0 0}, clip]{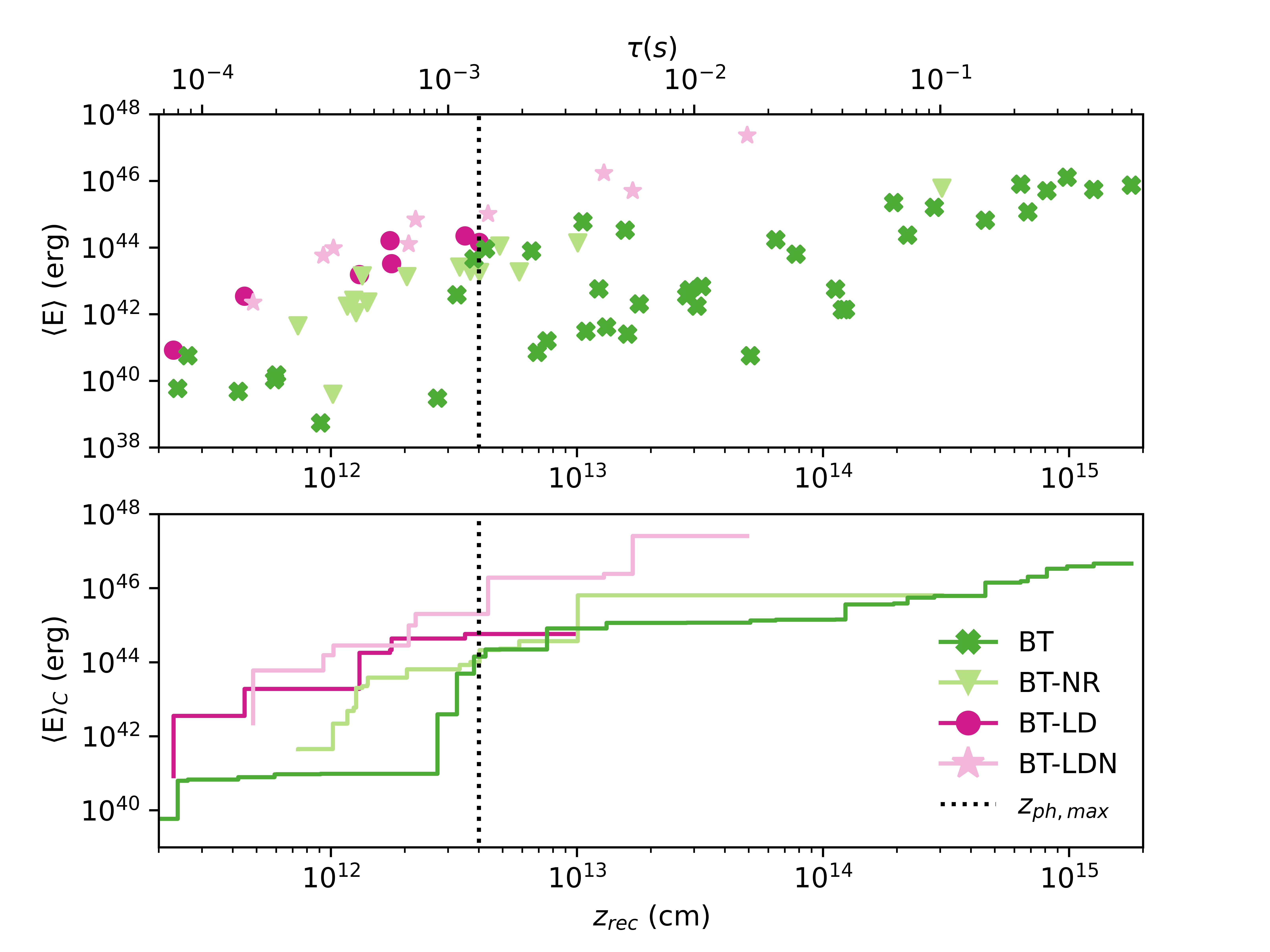}
\caption{Time-averaged energy $\langle E \rangle$ per stripe (top panel) and cumulative energy $\langle E \rangle_{\rm c}$ of all stripes (bottom panel) as a function of radial distance $z$ within the jet. 
The BT-LD model (see legend) dissipates all available energy within the estimated upper bound of the photospheric radius  $z_{\rm ph, max}$ (vertical line), while the remaining models dissipate only a small fraction ($\lesssim 1.6 \%$) of $\langle E \rangle_{\rm c}$. Here, $\langle E \rangle$ ($\langle E \rangle_{\rm c}$) appears to constitute a power-law as a function of radial distance and equivalently $\tau$. Fitting each model, the slope can be estimated as ($\langle E \rangle: 1.7$, $\langle E \rangle_{\rm c}: 1.8$) for the BT model, ($\langle E \rangle: 1.5$, $\langle E \rangle_{\rm c}: 1.7$) for BT-NR, ($\langle E \rangle: 2.5$, $\langle E \rangle_{\rm c}: 3$) for the BT-LD model, and ($\langle E \rangle: 2.2$, $\langle E \rangle_{\rm c}: 2.2$) for the BT-LDN model. }
\label{fig:energy_vs_zrec}
\end{figure}
%%%%%% FIGURE END %%%%%%%%

GRB jets can be opaque to electron scattering, i.e. if the optical depth is greater than unity, within the photospheric radius $z_{\rm ph}$ \citep{Giannios2012,Giannios2019}
\eqb
\label{eqn:photosphere_radius}
z_{\rm ph} \approx 4 \times 10^{12} \, L_{52} \, \Gamma_{\infty, 300}^{-1} \, \Gamma_{j, 100}^{-2} \, {\rm cm}.
\eqe
Here, $L$ is the jet's isotropic equivalent luminosity, $\Gamma_{\infty}$ is the jet's asymptotic Lorentz factor, and we have adopted the notation $Q_x = Q/10^x$ in cgs-units. 
Far within $z_{\rm ph}$, the radiated emission can be thermalized, leading to an apparent black-body spectrum from the photosphere \citep{giannios_spruit_2005}. Dissipation occurring close to the photosphere, $z=z_{\rm ph}$, leads to Comptonization of photons by hot electrons resulting in a high-energy power-law tail of the emission spectrum \citep{giannios2006}  while at $z \gg z_{\rm ph}$, the radiation is dominated by non-thermal synchrotron emission. 
The isotropic jet equivalent luminosity can be estimated as $L = L_{\rm j} / f_b$, where $L_{\rm j}$ is the intrinsic jet luminosity and $f_b = 1 - \cos \theta_{\rm j}$ is the beaming factor assuming a jet opening angle of $\theta_{\rm j}$. \citet{Christie2019} showed that the peak of $L_{\rm j}$ and time-averaged opening angle, $\langle \theta_{\rm j} \rangle$, were highly dependent upon the initial magnetic field configuration within the disk, spanning ranges of $L_{\rm j} \sim 10^{50} - 10^{52}$~erg~s$^{-1}$ and $\langle \theta_{\rm j} \rangle \sim 4.6^\circ - 13^\circ$. For the purely toroidal models presented here, we find similar jet peak powers in the range, $L_{\rm j} \sim 10^{50} - 10^{51}$~erg~s$^{-1}$. The corresponding isotropic equivalent luminosities are then estimated as $L \sim 10^{51} - 10^{53}$~erg~s$^{-1}$.
Adopting these and the fiducial values of $\Gamma_{\infty} \approx 300$ and $\Gamma_{\rm j} =100$, the photospheric radius lies within $z_{\rm ph} \sim 4 \times 10^{10}$~cm $ - 4 \times 10^{12}$~cm\footnote{For a more detail estimate of $z_{\rm ph}$ in the case of GRB 170817A, see \citep{meng2018}.} $ = 8 \times 10^{5} r_g - 8 \times 10^{7} \, r_g$.

In Fig.~\ref{fig:power_vs_dt}, these characteristic $z_{\rm ph}$-values are overplotted by vertical black lines. 
Within the estimated lower bound, we find only one stripe from the BT model, with $\tau \sim 3 \times 10^{-6} $~s, reconnects deep in the jet with a low $\langle P_{\Phi} \rangle$ of $\sim 7 \times 10^{41}$~erg~s$^{-1}$.
Stripes reconnecting within the upper estimated bound of $z_{\rm ph} \approx 4 \times 10^{12}$~cm, can have durations up to $\sim$ms timescales. The average power of these stripes can span $\sim 5$ orders of magnitude.
Stripes reconnecting outside our upper estimates of $z_{\rm ph}$ display a large spread of durations, ranging from $\sim$ms timescales to $\sim1$~s, and luminosities. 
Although a majority of these stripes are produced within the BT model, we find that the most luminous stripes, as produced from the BT-LDN model with $\tau \sim 10$~ms and $\langle P_{\Phi} \rangle \sim 10^{49}$~erg~s$^{-1}$, will reconnect outside the upper bound  of $z_{\rm ph}$.

However, in reconnection, we are not only interested in the power contained within the stripes, but also the amount of magnetic energy, estimated as $\langle E \rangle = \langle P_{\Phi} \rangle \, \tau$, available to be transferred to particle acceleration. 
More importantly, knowing the energy dissipation profile within the jet could assist in deciphering the origin of the prompt emission and the shape of the observed spectrum. 
In Fig.~\ref{fig:energy_vs_zrec}, we plot the time-average energy $\langle E \rangle$ within each stripe (top panel) and the cumulative energy $\langle E \rangle_{\rm c}$ of all dissipating stripes (bottom panel) as a function of radial distance within the jet. 
In both panels, we see that the average energy per stripe is larger for the BT-LD and BT-LDN models. 
To obtain an estimate for the energy dissipation profile within the jet, we can perform a power-law fit to both $\langle E \rangle \propto \tau^\alpha$ and $\langle E \rangle_{\rm c} \propto \tau^\beta$. 
Here, $\alpha$ ($\beta)$ is estimated as $\sim 1.7$ ($1.8$) for the BT model, $\sim 1.5$ ($1.7$) for BT-NR, $\sim 2.5$ ($3$) for BT-LD, and $\sim 2.2$ ($2.2$) for BT-LDN. We note that $\alpha \approx \beta$ for all of our models although we might have expected $\beta = \alpha + 1$ if $\langle E \rangle$ followed an exact power-law without any scatter. This apparent inconsistency arises precisely due to the scatter: the value of $\alpha$ is dominated by the average power of the jet, whereas the value of $\beta$ reflects the cumulative energetics of the jet and is dominated by the most energetic stripes (separated by quieter periods). Overall, $\beta$ is the more useful measure of the jet dissipation energetics. 
For the cumulative energy $\langle E \rangle_c$ within the jet,  
we see that all of the energy ($\sim 5.8 \times 10^{44}$~erg) for the BT-LD model will be dissipated at lower radii and within the estimated upper bound of the photospheric radius (see vertical line). 
The remaining three models, however, dissipate only a small fraction of the cumulative energy within $z_{\rm ph, max}$ with: the BT model dissipating $1.4\times10^{44}$~erg ($\sim 0.3 \%$), $10^{44}$~erg ($\sim 1.6 \%$) for the BT-NR model, and $2\times10^{45}$~erg ($\sim 0.8 \%$) for our BT-LDN model.  

\section{Discussion \& Conclusion}
\label{sec:discussion}

Here, we investigate the properties of naturally forming striped GRB jets as produced from 3D GRMHD simulations of post NS merger disks. 
Beginning with a purely toroidal magnetic field and slight variations within the micro (i.e. nuclear recombination, neutrino emission) and macro (i.e. more massive disk) properties of the disk (see Table~\ref{table:model_setups}), we find that all models undergo a similar dynamo-like action within the disk, producing both large scale poloidal magnetic flux (see Fig.~\ref{fig:flux_vs_t_zoom}) and jets of alternating polarity. 
In addition to finding that more massive disks produce stripes of larger power (Fig.~\ref{fig:power_vs_dt}), they accumulate larger amounts of a single-polarity magnetic flux on the BH, resulting in the cessation of striped jets. 
Although the distributions in the magnetic stripe duration $\tau$ (Fig.~\ref{fig:time_vs_dt}), power $\langle P_\Phi \rangle$ (Fig.~\ref{fig:power_vs_dt}), and energy $\langle E \rangle$ (Fig.~\ref{fig:energy_vs_zrec}) vary between all models, the robustness of striped-jet production signifies that it may be an inherent result of toroidally embedded, compact disks and are less governed by microphysical details.

Here, we used flux through the black hole as proxy for the jet power. For stripes with $\tau> 10^{-4}$~s this is a good approximation, but for shorter duration flips with $\tau\sim 10^{-5}$~s this is not as good of an approximation as the flips may not result in active jet phases. That being said, a large portion of the short duration stripes occurring before the formation of the jet and those which occur after reconnecting within the photosphere are not observable.

The power displayed within short GRB jets is highly variable, ranging on timescales from sub-ms to $\sim 1$~s \citep[observed in both simulations and observations][]{abbott2017a, abbott2017b,Christie2019, Gottlieb2020, oreshortgrb}. 
The dissipation of our observed stripes at a radial distance of $z_{\rm rec}$ within the jet corresponds to an observed variability on timescales of $t_v \simeq z_{\rm rec} / (\Gamma_j^2 c) \sim \tau / \beta_{\rm rec}$. 
Therefore, the most powerful stripes from our models, with $\tau$ ranging from $\sim  0.5 - 50$~ms (see Fig.~\ref{fig:power_vs_dt}), will display an observed variability ranging from $\sim 5$~ms to $\sim 0.5$~s, consistent with observations.
As compared with former theoretical models of striped jets, the energy dissipation profile found here is distinctly different. Some models \citep{Spruit2001,Drenkhahn2002,Drenkhahn_spruit_2002} calculate the energy dissipation profile assuming that the stripes are  injected with a unique, characteristic timescale. These studies included the bulk acceleration profile for the jet, whereas we assumed a consant Lorentz factor in the dissipation region. On the other hand, \citet{Giannios2019}, assumed an extended profile for the distribution of stripe lengthscales within the jet, finding a shallow cumulative energy profile $\langle E \rangle_{\rm c}\propto z^{1/3}$ up to a critical dissipation distance after which $\langle E \rangle_{\rm c}$ saturates at its asymptotic value. Moreover, they effectively associated the dominant stripe duration with timescales related to the inner edges of the disk, as might be expected in an extended, quasi-steady disk. Here, however, our disks are finite, narrow, and expand radially thereby introducing stripes of progressively longer duration within the jet. Our simulations point to having more magnetic flux associated with the expanding outer edge of the disk and, therefore, with progressively longer timescales. Moreover, we calculate the jet dissipation profile assuming a fiducial value for the jet Lorentz factor, $\Gamma_{\rm j} = 100$. This assumption has a strong effect on whether the dissipation in the jet takes place at high or small Thomson optical depths. For instance, for slower jets (e.g. $\Gamma_{\rm j} \sim 30$), the distance, $z_rec$, at which reconnection occurs (eqn.~\ref{eqn:z_rec}) decreases with increasing photospheric radius, $z_{\rm ph}$ (eqn.~\ref{eqn:photosphere_radius}), resulting in the majority of the stripes in all our models dissipating within $r_{\rm ph}$.

Our post-merger disk simulation results differ from those for radially extended disks of \citet{Liska_dynamo}, although both kinds of disks feature initially toroidal magnetic fields.
Namely, \citet{Liska_dynamo} found that although field reversals similarly occur within the disk, most of the generated poloidal magnetic flux loops are ejected as outflows from the disk: this results in a single magnetic loop dominating both the jet energetics and magnetic flux polarity.  
\citet{Christie2019}, argued that these differences were due to either: (i) the smaller post-merger disks being more tightly bound and less conducive to outflows and/or (ii) neutrino cooling producing even more tightly bound disks. 
We show that striped jets persist, albeit exhibiting different distributions in $\tau$ and $\langle P_\Phi\rangle$, across a range of microphysics (e.g., with vs without neutrino cooling), system properties (e.g., different merger remnant disk masses), and simulation numerical resolutions (see Table~\ref{table:model_setups}).
However, we note that the three new models presented here in comparison to \citet{Christie2019}, namely BT-NR, BT-LD, and BT-LDN, differ from the base BT model in accumulating sufficient magnetic flux by $t\sim 0.1$~s (see Fig.~\ref{fig:flux_vs_t_zoom}) such that the stripes are no longer produced within the jet at later times.     
It is useful to benchmark these results with future studies that include, in the context of post-merger disks, more realistic microphysics and initial conditions \citep[e.g. neutrino transport, realistic equation of state, more realistic post-merger initial conditions, and the inclusion of dynamical ejecta, see][]{Foucart2014,Foucart2019,miller2019}. 

%We use flux through the black hole as proxy for the jet power. For stripes with $\tau> 10^{-4}s$ this is a good approximation, but for shorter duration flips with $\tau\sim 10^{-5}$s this is not as good of an approximation as these flips may not result in active jet phases. That being said, a large portion of the short duration stripes occur before the formation of the jet, and those that occur after reconnect within the photosphere and thus are not observable.

\begin{acknowledgments}
EK acknowledges support by the National Science Foundation under Grant No. AST-1757792, a Research Experiences for Undergraduates (REU) grant award. IMC acknowledges support from the Fermi Guest Investigation grants 80NSSC18K1745 and 80NSSC20K0213. AT was supported by Fermi Cycle 14 Guest Investigator program 80NSSC22K0031, NASA grant 80NSSC18K0565, and by NSF grants
    AST-2107839, %new short GRB grant
    AST-1815304, %old short GRB grant
    AST-1911080, %accretion grant
    OAC-2031997, %Frontera travel grant
    AST-2206471. %TDE grant
    AT was partly supported by an NSF-BSF grant 2020747. DG acknowledges support from the NSF AST-2107802 and AST-2107806 grants.     This research used resources of the National Energy Research Scientific Computing Center, a DOE Office of Science User Facility supported by the Office of Science of the U.S. Department of Energy under Contract No. DE-AC02-05CH11231 using NERSC award NP-ERCAP0020543. The software used in this work was in part developed by the DOE NNSA-ASC OASCR Flash Center at the University of Chicago. Computations were performed at Carver, Hopper, and Edison (repositories m1186, m2058, m2401, and the scavenger queue).
\end{acknowledgments}


\bibliography{sample631}{}
\bibliographystyle{aasjournal}



\end{document}